\documentclass[11pt,a4paper]{article}
\pdfoutput=1

\pdfoutput=1
\usepackage{jcappub}
\usepackage[utf8]{inputenc}
\usepackage{graphicx}
\usepackage{color}
\usepackage{xfrac}
\usepackage{verbatim}
 \usepackage{mathtools}
\newcommand{\be}{\begin{equation}}
\newcommand{\ee}{\end{equation}}
\newcommand{\bea}{\begin{equation}\begin{aligned}}
\newcommand{\eea}{\end{aligned}\end{equation}}
\def\lsim{\mathrel{\raise.3ex\hbox{$<$\kern-.75em\lower1ex\hbox{$\sim$}}}}
\def\gsim{\mathrel{\raise.3ex\hbox{$>$\kern-.75em\lower1ex\hbox{$\sim$}}}}

\providecommand{\f}[2]{\frac{{#1}}{{#2}}}
\newcommand{\da}{\ensuremath{\dot{a}}}
\newcommand{\dda}{\ensuremath{\ddot{a}}}

\newcommand{\td}{\mathrm{d}}
\newcommand{\pd}{\partial}

\newcommand{\Tr}{{\rm Tr}}

\title{Quantum corrections to quartic inflation with a non-minimal coupling: metric vs. Palatini}

\author[a]{Tommi Markkanen,}
\author[b]{Tommi Tenkanen,}
\author[c]{Ville Vaskonen,}
\author[c,d]{and Hardi Veerm\"ae}

\affiliation[a]{Department of Physics, Imperial College London,\\Blackett Laboratory, London, SW7 2AZ, United Kingdom}
\affiliation[b]{Astronomy Unit, Queen Mary University of London, \\ Mile End Road, London, E1 4NS, United Kingdom}
\affiliation[c]{National Institute of Chemical Physics and Biophysics, \\ R\"avala 10, 10143 Tallinn, Estonia}   
\affiliation[d]{Theoretical Physics Department, \\ CERN, CH-1211 Geneva 23, Switzerland}
                            
\emailAdd{t.markkanen@imperial.ac.uk}
\emailAdd{t.tenkanen@qmul.ac.uk}
\emailAdd{ville.vaskonen@kbfi.ee}
\emailAdd{hardi.veermae@cern.ch}

\abstract{We study models of quartic inflation where the inflaton field $\phi$ is coupled non-minimally to gravity, $\xi \phi^2 R$, and perform a study of quantum corrections in curved space-time at one-loop level. We specifically focus on comparing results between the {\it metric} and {\it Palatini} theories of gravity. Transformation from the Jordan to the Einstein frame gives different results for the two formulations and by using an effective field theory expansion we derive the appropriate $\beta$-functions and the renormalisation group improved effective potentials in curved space for both cases in the Einstein frame. In particular, we show that in both formalisms the Einstein frame depends on the order of perturbation theory but that the flatness of the potential is unaltered by quantum corrections.}

\begin{document}
\begin{flushleft}
	\hfill		  IMPERIAL/TP/2017/TM/03 \\
	\hfill		  CERN-TH-2017-267
\end{flushleft}
\maketitle

\section{Introduction}

Many successful models of cosmic inflation exhibit a non-minimal coupling between matter fields and gravity~\cite{Ade:2015lrj}. One particularly appealing model is the one where the Standard Model (SM) Higgs boson acts as the inflaton field~\cite{Bezrukov:2007ep}, and where the non-minimal coupling between the Higgs field, $\Phi$, and gravity is of the form $\xi \Phi^\dagger\Phi R$, where $R$ is the Ricci scalar. While at tree-level the model predicts values for the spectral index of the primordial curvature power spectrum and tensor-to-scalar ratio which are in perfect agreement with observations of the Cosmic Microwave Background radiation spectrum (CMB)~\cite{Ade:2015lrj}, quantum corrections tend to make the scenario more complicated.

The first issue arises from the fact that for the best-fit values of SM parameters measured at the Large Hadron Collider (LHC), the SM potential is not stable up to the energy scales where inflation is to occur~\cite{Degrassi:2012ry,Buttazzo:2013uya,Bednyakov:2015sca}. Another issue is the claimed violation of perturbative unitarity at scales relevant for inflation~\cite{Burgess:2009ea,Barbon:2009ya,Barvinsky:2009ii,Bezrukov:2010jz,Burgess:2010zq,Hertzberg:2010dc}. Both of these issues have been addressed in a number of works during the recent years, for example in Refs.~\cite{Calmet:2013hia,Hamada:2014iga,Burgess:2014lza,Bezrukov:2014ipa,Hamada:2014wna,Bezrukov:2017dyv, Fumagalli:2016lls, Enckell:2016xse,Bezrukov:2014bra,Bezrukov:2014ina,Ballesteros:2015iua,Fumagalli:2017cdo, Kannike:2015apa}. Besides these matters being of significant theoretical interest, the quantum corrections can result also in other observable consequences such as primordial black holes (PBHs) which can provide a component of dark matter~\cite{Carr:2016drx,Carr:2017jsz}. Inflationary PBH production generally requires a large peak in the primordial power spectrum at small scales. This feature can be present for example in models of single field inflation with tuned potentials possessing inflection points or shallow local minima~\cite{Garcia-Bellido:2017mdw,Ezquiaga:2017fvi,Kannike:2017bxn, Germani:2017bcs, Gong:2017qlj, Motohashi:2017kbs, Pattison:2017mbe, Ballesteros:2017fsr}, see however Ref.~\cite{Bezrukov:2017dyv} which contested the results of Ref.~\cite{Ezquiaga:2017fvi}.
The PBH abundance is exponentially sensitive to the height of that peak if the overdense patch enters horizon during radiation dominance~\cite{Carr:1975qj}, thus even a small modifications to the potential could significantly alter the PBH abundance. Knowing the effect of quantum corrections is therefore important not only for determining whether the Higgs field can be responsible for inflation but also for other phenomena, such as the origin of dark matter.

The Higgs inflation is arguably the most minimal scenario where inflation can be realised. However, most authors have studied inflation only in the usual {\it metric} formalism, where one assumes not only a non-minimal coupling to gravity but also the Levi-Civita connection, which is uniquely determined by the space-time metric $g_{\mu\nu}$. At first, this might not look like an extra assumption, given that we are accustomed to using it in the context of the theory of General Relativity (GR), which we know to describe gravity very well not only at the distance scales within our Solar System but also at cosmological scales relevant for large scale structure formation. However, there is no \textit{a priori} reason to assume the Levi-Civita connection -- the connection $\Gamma$ can be treated as a free parameter. This is known as the \textit{Palatini} formalism.

In the context of GR the metric formalism coincides with the one of Palatini, as minimising the Einstein-Hilbert action with respect to the {connection} uniquely fixes it to be of the Levi-Civita form, $\Gamma=\Gamma(g_{\mu\nu})$. In more general models, however, especially in the ones involving matter fields that are non-minimally coupled to gravity, these two formalisms lead to two inherently different gravitational theories~\cite{Sotiriou:2008rp}. In particular, this means that inflationary models with non-minimal couplings to gravity cannot be characterised just by the scalar field potential, but also the fundamental gravitational degrees of freedom need to be specified, as was originally pointed out in Refs.~\cite{Bauer:2008zj,Bauer:2010jg}, and recently studied in~\cite{Rasanen:2017ivk,Tenkanen:2017jih,Racioppi:2017spw}.

The effect of quantum corrections on inflation has been studied extensively in the past, for recent works see for example \cite{Bilandzic:2007nb,Sloth:2006nu,Sloth:2006az,Enqvist:2013eua,Herranen:2015aja,Markkanen:2012rh,Herranen:2016xsy} and \cite{George:2013iia,George:2015nza,Bezrukov:2014ipa,Saltas:2015vsc,Bezrukov:2014bra,Cook:2014dga,Bezrukov:2009db} for Higgs inflation specifically. The effect of quantum corrections to models of Palatini inflation has been considered in~\cite{Rasanen:2017ivk,Racioppi:2017spw}. In the case of Palatini gravity, however, previous studies have assumed a specific form of the quantum corrections~\cite{Rasanen:2017ivk, Racioppi:2017spw}. Hence, there is an important question which has not been discussed in the literature before: calculating, from first principles, the effect of quantum corrections on inflation with a non-minimal coupling to gravity in the Palatini formalism. Furthermore, when quantum corrections are included, it remains to be verified that it is possible to support a sufficiently long period of inflation and obtain the correct, almost scale-invariant spectrum of perturbations. These questions we aim to explore in this paper, by using the specific example of quartic inflation as a representative case. 

The paper is organised as follows: In Sec.~\ref{model} we review inflation with a non-minimal coupling to gravity at tree-level in both metric and Palatini formalisms. In Sec.~\ref{quantumcorrections} we perform a general analysis of quantum corrections in curved space-time in both metric and Palatini formalisms, focusing for simplicity on the $\lambda\phi^4$ theory with a non-minimal coupling to gravity. In Sec.~\ref{results} we present our results and discuss the implications of quantum corrections on different models of inflation. We summarise our key conclusions in Sec.~\ref{conclusions}.

Our sign conventions are  $(+,+,+)$ according to the classification of \cite{Misner:1974qy}.

\section{Inflation with a non-minimal coupling to gravity}
\label{model}
We begin by comparing a generic model of cosmic inflation with a non-minimal coupling to gravity in the metric and Palatini cases. In both cases, the Jordan frame action reads
\be \label{nonminimal_action}
	S_J = \int d^4x \sqrt{-g}\left(\frac{1}{2}\left(M^2 + \xi\phi^2\right) g^{\mu\nu}R_{\mu\nu}(\Gamma) - \frac{1}{2} g^{\mu\nu}\partial_{\mu}\phi\partial_{\nu}\phi - V(\phi) \right) \,,
\ee
where $M$ is an arbitrary mass scale, $R_{\mu\nu}$ is the Ricci tensor, $\xi$ is a dimensionless coupling constant, $\phi$ is the inflaton field, $g_{\mu\nu}$ is the metric with $g$ its determinant, and $\Gamma$ is the connection. In the metric formulation $\Gamma=\bar{\Gamma}$, with
\be
	\bar{\Gamma}^\lambda_{\alpha\beta} = \frac{1}{2}g^{\lambda\rho}(\partial_\alpha g_{\beta\rho} + \partial_\beta g_{\rho\alpha} - \partial_\rho g_{\alpha\beta}) \,,
\ee
the Levi-Civita connection. In the Palatini formalism both $g_{\mu\nu}$ and $\Gamma$ are treated as independent variables. It is only assumed that the connection is torsion-free, $\Gamma^\lambda_{\alpha\beta}=\Gamma^\lambda_{\beta\alpha}$. In GR the equation of motion for the {connection} imposes $\Gamma=\bar{\Gamma}$ and hence renders the two formalisms equivalent. Despite this fact, for modified theories of gravity, especially with non-minimally coupled matter fields, this is generally not the case~\cite{Sotiriou:2008rp}. Because of that, models of cosmic inflation in the Palatini and metric formulations are intrinsically different, as originally discussed in Ref.~\cite{Bauer:2008zj}. The Palatini formulation swaps one theory for another~\cite{Iglesias:2007nv}, and this is exactly what we will see: the Einstein frame potentials (where the non-minimal couplings vanish) for the two formulations will be different. As we will show, this also means that the quantum corrections and their possible consequences are different for these two cases.

The non-minimal coupling in the Jordan frame action \eqref{nonminimal_action} can be removed by the following conformal transformation,
\be \label{Omega}
	g_{\mu\nu} \to \Omega(\phi)^{-1}g_{\mu\nu}, \hspace{1cm} \Omega(\phi)\equiv \frac{M^2+\xi \phi^2}{M_{\rm P}^2} \,,
\ee
where $M_{\rm P}$ is the reduced Planck mass. Notice that, in general, the scale $M$ may differ from $M_{\rm P}$. With this, we obtain the Einstein frame action
\be \label{einsteinframe}
	S_E = \int d^4x \sqrt{-g}\left(\frac{1}{2}M_{\rm P}^2g^{\mu\nu}R_{\mu\nu}(\Gamma)-\frac{M^2+\xi\phi^2+6f\xi^2\phi^2}{2\Omega(\phi)\left(M^2+\xi\phi^2\right)}\, g^{\mu\nu} \partial_{\mu}\phi\partial_{\nu}\phi - \frac{V(\phi)}{\Omega(\phi)^2} \right) \,,
\ee
where $f=1$ ($f=0$) in the metric (Palatini) case.

Because in the Einstein frame the connection appears only in the Einstein-Hilbert term, the equations of motion guarantee that the Ricci tensor can be expressed as a function of the Levi-Civita connection.\footnote{This is generally not the case if the Lagrangian contains fermions, because fermionic kinetic terms depend on the spin connection. This contribution may, however, be compensated by including suitable contact terms quadratic in torsion~\cite{Deser:1976ay} or by imposing $\Gamma=\bar{\Gamma}$ with Lagrange multipliers~\cite{Hehl:1994ue,Hehl1978}.} Thus we can impose the on-shell relation $\bar{\Gamma} = \Gamma$. The non-minimal Palatini formalism is hence equivalent to using a Jordan frame action \eqref{nonminimal_action} with the Einstein frame Levi-Civita connection. With a suitable field redefinition $\phi = \phi(\chi)$, determined by
\be \label{chi}
	\frac{d\phi}{d\chi} = \sqrt{\frac{\Omega(\phi)\left(M^2+\xi\phi^2\right)}{M^2+\xi\phi^2+6f\xi^2\phi^2}} \,,
\ee
the kinetic terms can be brought to the canonical form. The action~\eqref{einsteinframe} then becomes
\be \label{EframeS}
	S_{\rm E} = \int d^4x \sqrt{-g}\bigg(\frac{1}{2}M_{\rm P}^2R -\frac{1}{2}{\partial}_{\mu}\chi{\partial}^{\mu}\chi - U(\chi)  \bigg) \,,
\ee
where $U(\chi) =V(\phi(\chi))/ \Omega^{2}(\phi(\chi))$ and $R = g^{\mu\nu}R_{\mu\nu}(\bar{\Gamma})$.  

From now on we will focus on a specific form of the Jordan frame potential, namely, one with the leading contribution as
{
\be\label{phipotential}
	V(\phi)=\frac{\lambda}{4} \phi^4 \,.
\ee
In this model inflation takes place at large field values for which the Einstein frame potential develops a plateau if $\Omega(\phi) \propto \phi^2$. For \eqref{Omega} it is realised when $\phi\gg M/ \sqrt{\xi}$. The lower order terms in the potential may induce a vacuum expectation value for the field $\phi$, which we assume to be much smaller than the scale $M/\sqrt{\xi}$. In terms of the canonical field, the large field Einstein frame potential reads}\footnote{We use the asymptotic boundary condition $\phi(\chi \to \infty) = \frac{M}{2\sqrt{\xi}} \exp\left( c \chi/M_{\rm P} \right)$ when solving \eqref{chi}, where $c = 1/\sqrt{6}$ in the metric and $c = \sqrt{\xi}$ in the Palatini case.}
\bea \label{chipotential}
	U(\chi) &\simeq 
\begin{cases}	
	\frac{\lambda M_{\rm P}^4}{4\xi^2}\bigg(1-8\exp\bigg(-\sqrt{\frac{2}{3}}\frac{\chi}{M_{\rm P}} \bigg) \bigg)  & \quad \mathrm{metric} ,\\
	\frac{\lambda M_{\rm P}^4}{4\xi^2}\left(1-8\exp\left(-\frac{2\sqrt{\xi}\chi}{M_{\rm P}} \right)\right) & \quad \mathrm{Palatini} , 
\end{cases}
\eea
in the metric and Palatini cases, respectively.  At high field values, that is for $\chi\gg M_{\rm P}$ in the metric case, or $\chi\gg M_{\rm P}/\sqrt{\xi}$ in the Palatini case, the potential tends to a constant exponentially fast.

In slow-roll approximation the inflationary dynamics is characterised by the slow-roll parameters 
\be
\label{SRparameters}
	\epsilon \equiv \frac{1}{2}M_{\rm P}^2 \left(\frac{1}{U}\frac{{\rm d}U}{{\rm d}\chi}\right)^2 \,, \quad
	\eta \equiv M_{\rm P}^2 \frac{1}{U}\frac{{\rm d}^2U}{{\rm d}\chi^2} \,, \quad
\ee
and the total number of e-folds during inflation
\be \label{Ndef}
	N = \frac{1}{M_{\rm P}^2} \int_{\chi_f}^{\chi_i} {\rm d}\chi \, U \left(\frac{{\rm d}U}{{\rm d} \chi}\right)^{-1},
\ee
where the field value at the end of inflation, $\chi_f$, is defined via\footnote{In the Palatini formalism $|\eta|=1$ occurs actually earlier than $\epsilon=1$. However, the moments when $|\eta|=1$ and $\epsilon=1$ are separated only by approximately one $e$-fold, so as a good estimate one can use the slow-roll approximation down to $\epsilon=1$.} $\epsilon(\chi_f) = 1$. 

The large $N$ expansion of the spectral index, $n_s \simeq 1 - 2\epsilon + \partial_N\ln\epsilon$, and tensor-to-scalar ratio, $r\simeq 16\epsilon$, reads 
\bea \label{nsr}
	n_s(\chi_i) &\simeq 1 - \frac{2}{N}  \, \ \quad \mbox{metric and Palatini} ,\\
	r(\chi_i) &\simeq 
	\begin{cases}
		\frac{12}{N^2}  & \quad \mathrm{metric} ,\\
		\frac{2}{\xi N^2} & \quad \mathrm{Palatini} . \\
	\end{cases}
\eea
Knowing the amplitude of  the curvature power spectrum, $U(\chi_i)/\epsilon(\chi_i) = (0.027M_{\rm P})^4$~\cite{Lyth:1998xn,Ade:2015xua}, relates the non-minimal coupling to the number of required e-folds and the quartic self-coupling strength of the inflaton as
\be \label{xicondition}
	\xi \simeq 
	\begin{cases}
		790 N\sqrt{\lambda} & \quad \mathrm{metric} ,\\
		3.8\times10^6 N^2\lambda & \quad \mathrm{Palatini} . \\
	\end{cases}
\ee
For instance, $\lambda\simeq 0.1$ and $N=60$ require $\xi=\mathcal{O}(10^4)$ in the metric case and $\xi=\mathcal{O}(10^9)$ in the Palatini case. 

The numerical values for $N=60$ are $n_s=0.968$ and $r=3\times10^{-3}$ (metric), $r=4\times10^{-14}/\lambda$ (Palatini). Both results are in good agreement with the state-of-the-art results $n_s=0.9677\pm 0.0060$ ($68\%$ confidence level) and {$r<0.12$ ($95\%$ confidence level) \cite{Ade:2015lrj,Array:2015xqh}}. In the metric case the predicted tensor-to-scalar ratio is within the reach of planned future experiments such as LiteBIRD~\cite{Matsumura:2013aja}, but unless $\xi\lsim1$ (in which case there would also be corrections to the above approximations), the tensor-to-scalar ratio in the Palatini case is too small to be within the reach of any future experiment.

\section{Quantum corrections in curved space}
\label{quantumcorrections}

We now turn to discuss quantum corrections in curved space-time. First we derive the effective potential in curved space, and then consider its renormalisation group improvement. For related work discussing similar issues see for example Refs.~\cite{Hu:1984js,Barvinsky:1993zg,Elizalde:1993ee,Elizalde:1993ew,Kirsten:1993jn,Herranen:2015aja,Bounakis:2017fkv,George:2012xs,Elizalde:2015nya,Herranen:2014cua,Codello:2015oqa}.

We will consider only the contribution from the quantum fluctuations of the scalar field. We will be interested in the "leading log" results, so we only include logarithmic contributions in the ultraviolet limit. These are the only relevant contributions for determining the specific form of the operators generated by quantum corrections and their renormalisation group running, which will be addressed in the next section. Such approximations are often made in works investigating similar issues~\cite{Sloth:2006nu,Sloth:2006az,Enqvist:2013eua,Markkanen:2012rh,Herranen:2016xsy,George:2013iia,Bezrukov:2014ipa,Cook:2014dga,Bezrukov:2009db}. In this section we do not specify the Jordan frame potential, making the derivation of this section applicable to cases beyond the quartic form (\ref{phipotential}).

Having brought the Jordan frame action \eqref{nonminimal_action} into the canonical form \eqref{EframeS}, let us now focus on the minimally coupled scalar field in a curved background. It is described by the action 
\be \label{eq:actS}
	S_{\chi} = \int \td^4x\,\sqrt{-g}\left(-\f{1}{2}\partial_\mu\chi\partial^\mu\chi-U(\chi)\right) \,.
\ee
For a simple scalar field in the one-loop approximation the quantum corrected energy-momentum tensor and  equation of motion can be derived without a need for more sophisticated approaches. We start by splitting the field into its expectation value and a quantized fluctuation, $\chi \equiv \langle\chi\rangle + \hat{\chi}$. To one-loop order the equation of motion for the mean field $\langle\chi\rangle$ and the fluctuation $\hat{\chi}$ can be derived by first expanding the action up to the quadratic order
\bea \label{eq:expS}
	S_{\chi} 
	&= -\int \td ^4 x\sqrt{-g}~ \left(\f{1}{2}\partial_\mu\chi\partial^\mu\chi +U(\chi) \right) -\int \td ^4 x\sqrt{-g}~\f{1}{2} \hat{\chi} \left(-\Box+m^2(\chi) \right)\hat{\chi} + \dots \,,\\
	&=S_{\chi}^{(0)}+S_{\chi}^{(1)}+\dots
\eea
where $\chi \equiv \langle\chi\rangle$ is the classical field and we defined the effective mass\footnote{In case of multiple fields the mass matrix reads $(m^{2})_{ij} \equiv \pd^2 U/ \pd \chi^i \pd \chi^j$. After a change of variables $\chi^i \mapsto \phi^i$ the mass matrix can be written as 
\be \label{def:M2_gen} \nonumber
	(m^{2})_{ij} = \frac{\pd \phi^k} {\pd \chi^i} \frac{\pd \phi^l} {\pd \chi^j} \left(\frac{\pd^2 U}{ \pd \phi^k \pd \phi^l} - \Gamma^{m}{}_{kl}  \frac{\pd U}{ \pd \phi^m} \right) \equiv \gamma^{ik} \nabla_k \nabla_j U,
\ee
where  $\nabla_i$ is the covariant derivative, $\Gamma^{k}{}_{ij} = (\pd \phi^k/\pd \chi^l) \, (\pd \chi^l / \pd \phi^i \pd \phi^j)$ are the Christoffel symbols corresponding to the field space metric $\gamma^{ij} = \delta^{lm} (\pd \phi^i/\pd \chi^l) \, (\pd \phi^j/\pd \chi^m) $. The mass matrix $(m^{2})_{ij}$ is defined for a specific choice of variables -- the canonical variables. The generalisation to curved field spaces, that is to non-canonical kinetic terms with general $\gamma^{ij}$, is given by the covariant mass matrix $(\bar m^2)^{i}_{j} \equiv \gamma^{ik} \nabla_k \nabla_j U$~\cite{George:2013iia}. Although $(\bar m^2)^{i}_{j}$ depends on the choice of variables, $\Tr(f(\bar m^2))$ does not ($f$ is an arbitrary function). Moreover, $\Tr(f(\bar m^2)) =  \Tr(f(m^2))$.}
\be
	m^2(\chi) = U''(\chi)\,.
\ee
The linear term is omitted as it is higher order: its expectation value vanishes at all orders since $\langle\hat{\chi}\rangle = 0$ by construction. 

{To find the effect of quantum fluctuations we need to solve the one-loop equation of motion for the quantum field
\be	\label{eq:eom3}
	\left(-\Box + m^2(\chi)\right)\hat{\chi} = 0 \,,
\ee	
In order to derive the leading running logs in curved space we define a properly normalised ansatz that can be used for solving the UV behaviour in any background, which resembles very much the adiabatic mode frequently used in the context of quantum field theory in curved space and in particular the technique of adiabatic subtraction \cite{Parker:1974qw,Fulling:1974pu}. However in contrast to an adiabatic expansion, we emphasise that at no point will we expand around small $H$, as in general this is not allowed during slow-roll inflation. 

Following the steps and notation of \cite{Markkanen:2013nwa}, the one-loop modes in cosmic time $\td s^2=-\td t^2+a^2 \td \mathbf{x}^2$ can be defined as}
\be\label{eq:adsol3}
	\hat{\chi}	=\int \td^{n-1}\mathbf{k}\, e^{i\mathbf{k\cdot\mathbf{x}}}\left(\hat{a}_\mathbf{k}^{\phantom{\dagger}}f^{\phantom{\dagger}}_{k}(t)+\hat{a}_{-\mathbf{k}}^\dagger f^*_k(t)\right)\,, \quad 
	{f}_{k}(t)	=\f{e^{-i\int^{t}W \td t'}}{\sqrt{2W(2\pi a)^{n-1}}}\,, 
\ee
where we have analytically continued the spacetime dimensions to $n$ in anticipation of dimensional regularisation. After inserting \eqref{eq:adsol3} into the (one-loop) equation of motion \eqref{eq:eom3} for the fluctuation one gets the following equation for $W$,
\be \label{eq:W}
	W^2 = \frac{\mathbf{k}^2}{a^2} + m^2(\chi)  + \f{(n-1)(3-n)}{4} \bigg(\f{\da}{a}\bigg)^2 - \f{n-1}{2}\f{\dda}{a} + \f{3\dot{W}^2}{4W^2} - \f{\ddot{W}}{2W}\,.
\ee
We can now solve for $W$ from~\eqref{eq:W} iteratively as an expansion in terms of large $k/a$. The first few orders may be written as
\be \label{sol:W}
	W=\sqrt{\f{\mathbf{k}^2}{a^2} + m^2(\chi) - \frac{n-2}{4 (n-1)}R+ \mathcal{O}(k^{-2})} \,.
\ee
In order to obtain all the generated divergences -- and hence the running logarithms -- the $\mathcal{O}(k^{-2})$ corrections are not required.

From now on, for simplicity, we will only consider the quantum corrections in pure de Sitter space. As the quantum contribution is expected to be small, neglecting the slow-roll corrections to the quantum corrections introduces a negligible error.

\subsection{Effective energy density in curved space}
\label{sec:em}
The one-loop contribution to the energy-momentum tensor is by definition
\be \label{eq:renomE}
	\hat{T}_{\mu\nu}^{(1)} 
	= -\f{2}{\sqrt{-g}}\f{\delta S_{\chi}^{(1)}}{\delta g^{\mu\nu}} \,.
\ee
The results \eqref{eq:adsol3},  \eqref{sol:W}  from the previous section then yield the energy density
\bea \label{eq:T00}
	&\langle\hat{T}_{00}^{(1)}\rangle 
	= \f{1}{2}\int {d^{n-1} {k}}\left(\vert\dot{f}_{k}\vert^2+\bigg( {\f{\mathbf{k}^2}{a^2}+m^2(\chi)}\bigg)\vert f_{k}\vert^2\right) \\ 
	&= \int  \f{d^{n-1} {k}}{(2\pi a)^{n-1}W}\left(\frac{W^2}{4} + \f{\mathbf{k}^2}{4a^2}+\frac{H^2 (n-1)^2 + 4m^2(\chi)}{16} + H(n-1)\frac{ \dot{W}}{8 W}+\frac{\dot{W}^2}{16 W^2}\right) \,,
\eea
with the understanding that only the divergences and the running logarithms are to be included in the end result. After some algebra we obtain the unrenormalised result
\be \label{eq:toor}
	\langle\hat{T}_{00}^{(1)}\rangle=m^2(\chi)\f{m^2(\chi)-2H^2}{64\pi^2}\log\bigg\vert\f{m^2(\chi)-2H^2}{\tilde{\mu}^2}\bigg\vert+\dots \,,
\ee
where the ellipses represent finite terms. The divergences have been included in the definition of the renormalisation scale
\be
	\log\tilde{\mu}=\log{\mu}-(n-4)^{-1}\,,
\ee
where $\mu$ is finite and arbitrary.

Evidently the quantum corrections introduce a term mixing the curvature and the potential despite the fact that we initially performed a Weyl scaling to remove any such terms from our tree-level Lagrangian in Sec.~\ref{model}. Specifically, in order to remove the divergences generated by the loop correction we need to introduce the following counter terms 
\be \label{eq:cts}
	S_{\chi} \rightarrow S_{\chi}+\delta S,\quad \text{where}\quad \delta S= -\int d^{n}x\,\sqrt{-g} \left(\f{\delta c_\xi}{2}m^2(\chi)R+\delta c_\lambda m^4(\chi)\right) \,.
\ee
and, for constant field configurations, the counter terms for the energy density are therefore
\be
	\delta T_{00} = \delta c_\xi G_{00}m^2(\chi)+\delta c_\lambda m^4(\chi)\,,
\ee
where $G_{00}$ is the $00$ component of the Einstein tensor. Thus, with the choices
\be \label{eq:divct}
	\delta c_\xi=\f{1}{48\pi^2}\f{1}{n-4} \,,\qquad \delta c_\lambda=-\f{1}{32\pi^2}\f{1}{n-4}\,,
\ee
we can simply set $\tilde{\mu}\rightarrow\mu$ in~\eqref{eq:toor}.

\subsection{Effective field equation in curved space}

\label{sec:eom}
Variation of the action~\eqref{eq:expS} with respect to the field leads to the quantum corrected equation of motion
\be \label{eq:eom2}
	-\Box \chi + U'(\chi) + \f{1}{2}U'''(\chi)\langle\hat{\chi}^2\rangle = 0 \,.
\ee
For a constant mean field $\chi$ the quantum corrected equation of motion~\eqref{eq:eom2} reduces to finding the minimum of the effective potential, whose one-loop approximation reads
\be \label{eq:effM}
	U_{\rm eff}'(\chi)\equiv \frac{\td}{\td\chi}\left(U^{(0)}(\chi)+U^{(1)}(\chi)\right)= U'(\chi) + \f{1}{2}U'''(\chi)\langle\hat{\chi}^2\rangle = 0 \,.
\ee

We can then proceed as in the previous section and calculate the one-loop contribution as
\bea \label{eq:varia}
	\frac{\td}{\td\chi}U^{(1)}(\chi)
	&=\f{U'''(\chi)}{2}\langle \hat{\chi}^2\rangle = \f{U'''(\chi)}{2}\int \f{\td^{n-1} k}{2(2\pi a)^{n-1}W} \\
	&=\f{U'''(\chi)}{2}\f{m^2(\chi)-2H^2}{16\pi^2}\log\bigg\vert\f{m^2(\chi)-2H^2}{\tilde{\mu}^2}\bigg\vert+\cdots .
\eea
By using the counter term action we introduced in~\eqref{eq:cts} one generates counter-terms for the equation of motion as 
\be
	\delta U'(\chi)=\f{\delta  c_\xi}{2}U'''(\chi)R+2\delta  c_\lambda U'''(\chi)m^2(\chi) \,,
\ee
from which it is easy to see that the counter terms $\delta c_\xi$ and  $\delta c_\lambda$ previously derived for the energy density in~\eqref{eq:divct} also renormalise the equations of motion, formally as $\tilde{\mu}\rightarrow\mu$. This is of course demanded by consistency, but nonetheless, as we will discuss in the next section, it leads to an important conclusion regarding the form of the quantum corrections that is not widely appreciated in literature.

We note that a very similar approach for deriving the quantum corrections was used in~\cite{Kohri:2016qqv} in the context of the electroweak vacuum instability during inflation.

\subsection{Form of the one-loop corrections}

The derivation of both the effective energy density Sec.~\ref{sec:em} and the effective field equations in Sec.~\ref{sec:eom} imply that operators of the form
\be \label{eq:genop}
	m^2(\chi)R\quad\text{and}\quad m^4(\chi) \,,
\ee
are generated by loop corrections despite their absence at tree-level. The results of the previous two sections correspond to the renormalised effective action
\begin{align} \label{eq:seff}
	S_{\chi}+\delta S
	&= \int \td^4x\,\sqrt{-g}\left(-\f{1}{2}\partial_\mu\chi\partial^\mu\chi-U_{\rm eff}(\chi)\right) \\
	&\equiv \int \td^4x\,\sqrt{-g}\left(-\f{1}{2}\partial_\mu\chi\partial^\mu\chi-U(\chi) {-} m^2(\chi)\f{m^2(\chi)-\frac{R}{3}}{64\pi^2}\log\left|\f{m^2(\chi)-\frac{R}{6}}{{\mu}^2}\right| \right) \,.\nonumber
\end{align}
It is straightforward to show by varying the above quantum correction and including only the leading logarithms that it reproduces the quantum corrections for the energy density and equation of motion for $\chi$, given respectively in~\eqref{eq:toor} and~\eqref{eq:varia}. The effective action \eqref{eq:seff} is consistent with results in earlier works using different approaches, see for example \cite{Herranen:2014cua,Kirsten:1993jn} and references therein.  

We stress that for a generic theory of inflation $R\gg m^2(\chi)$. Neglecting curvature in the loop corrections thus neglects the dominant quantum contribution and therefore approximating quantum corrections via flat space field theory is in many cases not sufficient.

In addition, we emphasise the unavoidable emergence of the $m^2(\chi)R$ term in both the metric and Palatini formulations. It is significant since if one were to remove this contribution via a tree-level Weyl scaling, it re-surfaces at one-loop order. This term cannot be removed from (\ref{eq:seff}) by making use of the tree-level Friedmann equation and expressing $R$ in terms of the field $\chi$, as this would lead to inconsistencies in the field equations. 
We can conclude that the non-minimally coupled form of the tree-level action will unavoidably be spoiled by the one-loop corrections. Maintaining a minimally coupled effective action requires an additional Weyl scaling that removes the quantum generated pieces mixing inflaton and curvature. Furthermore, higher orders in the loop expansion presumably generate additional non-minimal terms. The Einstein frame therefore depends on the order of perturbation theory.

The fact that the operators~\eqref{eq:genop} are included by the quantum corrections implies that they will be generated via renormalisation group running. This then means that there will be non-trivial $\beta$-functions associated with them. The $\beta$-functions are universal and hence independent of the used regulator. Their generation and existence is therefore unavoidable, which we will address next.

\subsection{Renormalisation group improvement and $\beta$-functions}

Now we may proceed to the main new results of this work, which come by calculating the renormalisation group improved effective potential in curved space both in the metric and Palatini formulations. 

The non-minimally coupled quartic inflation corresponds to exponential Einstein frame potentials given in~\eqref{chipotential}. Eq.~\eqref{eq:seff} shows that theories with such potentials are inherently non-renormalisable: in order to cancel the divergences generated by quantum corrections it is necessary to introduce operators not present at tree-level. Nevertheless, the Einstein frame effective potentials~\eqref{chipotential}, and hence the quantum corrections, contain an exponentially small parameter at the large field limit, which provides a natural small quantity one may use for an expansion in the effective theory sense. 

We assume that the theory is indeed well-defined as an effective field theory, as is often done~\cite{George:2013iia,Bezrukov:2014ipa,George:2015nza}. In that case all successive quantum corrections are appropriately suppressed and a valid expansion may be formed. Up to some given order of truncation, we can then renormalise our model order by order as for a renormalisable theory. The leading contribution is generated from the one-loop effective action~\eqref{eq:seff}. More precisely, we expand the tree-level potential in the small dimensionless quantity $\delta$ and the curvature scale $\propto H\ll M_{\rm P}$ as
\be \label{eq:efft}
	\bar{U}(\chi) =  c_0 M_{\rm P}^4 +c_1 R M_{\rm P}^2 + c_2 M_{\rm P}^4 \delta + c_3 M_{\rm P}^4  \delta^2  + \xi_{E} R \delta  \, M_{\rm P}^2 +\cdots\,,
\ee
where $\delta$ is defined as
\be \label{cts}
	\delta
	=\begin{dcases}
		\exp\bigg(-\sqrt{\frac{2}{3}}\frac{\chi}{M_{\rm P}} \bigg) & \quad \mathrm{metric} \,,\\
		\exp\left(-\frac{2\sqrt{\xi}\chi}{M_{\rm P}} \right) & \quad \mathrm{Palatini} \,. \\
	\end{dcases} 
\ee
Note that in terms of the Jordan frame field $\delta = M^2/(4\xi\phi^2)$. In the end of the calculation the free parameters in~\eqref{eq:efft} will be matched to the physical ones given by~\eqref{chipotential}.

As the renormalisation scale $\mu$ is arbitrary, the physics should not depend on it. The effective potential thus has to satisfy 
\be \label{eq:rg}
\frac{\td U_{\rm eff}(\chi)}{\td\mu}=0\,.
\ee
This requirement leads to the well-known Callan-Symanzik (CS) equations for the effective potential~\cite{Ford:1992mv}
\be \label{eq:CS}
\left(\mu\f{\partial}{\partial\mu}+\beta_{c_i}\f{\partial}{\partial{c_i}}-\gamma \chi\f{\partial}{\partial\chi}\right) U_{\rm eff}(\chi)=0\,, \qquad\beta_{c_i}\equiv \mu\f{\partial c_i}{\partial\mu} \,,
\ee
where the $c_i$ stands for all the parameters of the action (including $\xi_E$)  with summation over the repeated index $i$ assumed. Using the CS equation for solving for the $\beta$-functions is a simple exercise. Although, in general, it requires the field anomalous dimension $\gamma$ as an input, for our theory one has $\gamma=0$ at one-loop~\cite{George:2013iia}. Hence the effective potential~\eqref{eq:seff} implies the following one-loop CS equation
\be
\beta_{c_i} \frac{\partial}{\partial{c_i}} \bar{U}(\chi)= -\mu\frac{\partial}{\partial\mu } \left( m^2(\chi) \frac{m^2(\chi)-\frac{R}{3}}{64\pi^2}\log\left|\frac{m^2(\chi)-\frac{R}{6}}{{\mu}^2}\right| \right)\,,
\ee
which by using~\eqref{eq:efft} leads to $\beta_{c_0}=\beta_{c_1}=\beta_{c_2}=0$ and
\be \label{eq:betas}
	\beta_{\xi_{E}}=
	\begin{dcases}
		\f{\lambda}{72\pi^2\xi^2} & \quad \mathrm{metric}\,,\\
		\f{\lambda}{12\pi^2\xi}& \quad \mathrm{Palatini} \,, \\
	\end{dcases} 
	\qquad 
	\beta_{c_3}=
	\begin{dcases}
		\f{\lambda^2}{18\pi^2\xi^4} & \quad \mathrm{metric}\,,\\
		\f{2\lambda^2}{\pi^2\xi^2}& \quad \mathrm{Palatini} \,. \\
	\end{dcases}
\ee
In~\eqref{eq:betas} the $\lambda$ and $\xi$ couplings are the ones introduced in the Jordan frame and are constant. This is a manifestation of the fact that the quantum correction (\ref{eq:seff}) is entirely made of operators not present at tree-level (\ref{chipotential}) or, in the parametrisation (\ref{eq:efft}), the $\beta$-functions for $c_0$ and $c_2$ vanish. 

The leading quantum correction can from Eq.~\eqref{eq:efft} be seen to come from $\xi_E$ which can be solved to run as\footnote{Alternatively, the running can be calculated directly from the counter term~\eqref{eq:divct}. 
}
\be \label{eq:runx}
	\xi_E(\mu)=
	\begin{dcases}
		\f{\lambda}{72\pi^2\xi^2}\log\bigg(\f{\mu}{\mu_0}\bigg)  & \quad \mathrm{metric} ,\\
		\f{\lambda}{12\pi^2\xi}\log\bigg(\f{\mu}{\mu_0}\bigg) & \quad \mathrm{Palatini} , \\
	\end{dcases}
\ee
where we have implemented the choice $\xi_E(\mu_0)=0$.

Finally, by using the standard approach for renormalisation group improvement~\cite{Ford:1992mv}, we optimise our loop expansion by exploiting the $\mu$-independence of the result and choose the scale $\mu$ in such a way that that the logarithm in the loop correction of~\eqref{eq:seff} vanishes. So, we make the choice 
\be \label{scale}
	\mu^2=|m^2(\chi)-R/6|\approx 2H^2\,,
\ee
which allows us to write the one-loop RG improved effective potential in the large field regime to the lowest order in the expansion in $\delta$ in a very simple form
\be \label{RGpot0}
	U_{\rm eff}(\chi)= \frac{\lambda M_{\rm P}^4}{4\xi^2}\left(1-8\delta(\chi)\right) 
	+ 	\xi_E(\mu)R M_{\rm P}^2 \delta(\chi)+\dots
\ee
Again, the conformal transformation 
\be
	g_{\mu\nu} \to \tilde\Omega(\chi)^{-1}g_{\mu\nu}, \hspace{1cm} \tilde\Omega(\chi)\equiv 1-2\xi_E(\mu) \delta(\chi)\,,
\ee
removes the 1-loop induced non-minimal coupling. To the lowest order the kinetic term of $\chi$ remains intact, and the Einstein frame effective potential is $U_{{\rm eff},E}(\chi) = U_{\rm eff}(\chi)/\tilde\Omega(\chi)^2$, which by expanding in $\delta\ll1$ and $\xi_E\ll1$ gives\be \label{RGpot}
	U_{{\rm eff},E}(\chi) =
	\begin{dcases}
	\frac{\lambda M_{\rm P}^4}{4\xi^2}\left(1-8\left(1-\frac{\xi_E(\mu)}{2}\right)\exp\left(-\sqrt{\frac{2}{3}}\frac{\chi}{M_{\rm P}} \right) \right)+\dots  & \mathrm{metric} \,,\\
	\frac{\lambda M_{\rm P}^4}{4\xi^2}\left(1-8\left(1-\frac{\xi_E(\mu)}{2}\right)\exp\left(-\frac{2\sqrt{\xi}\chi}{M_{\rm P}} \right)\right)+\dots & \mathrm{Palatini} \,.
	\end{dcases}
\ee
We emphasise the scale $\mu$ in~\eqref{eq:runx} is to be chosen as~\eqref{scale}. Note that the leading order correction can also be obtained by simply plugging the tree level Friedmann equation $R=4U/M_{\rm P}^2$ into~\eqref{RGpot0}.

To the best of our knowledge the curved space RG improved effective potential~\eqref{RGpot} has not previously been calculated in the context of inflation with exponential potentials in particular in Higgs inflation, in neither the metric nor Palatini frameworks.

\section{Results}
\label{results}

An important feature of the quantum corrections that is sometimes underestimated in the literature is the generation of a non-minimal coupling for the tree-level Einstein frame action. In fact, during inflation this gives rise to the dominant quantum correction as is explicitly seen in the RG improved potential \eqref{RGpot}. The Jordan frame couplings $\xi$ and $\lambda$, on the other hand, do not run at one-loop.

The running of the regenerated non-minimal coupling \eqref{eq:runx} seems to indicate that the quantum corrections in the metric case seem to be suppressed by a factor of $1/\xi$ when compared to the metric case. This does not hold during inflation, however, and the corrections to inflationary observables are of the same order in both cases. The reason for this is that the scale of quantum corrections can be directly related to the observed magnitude of inflationary perturbations, which imply $U/\epsilon = (0.027M_{\rm P})^4$. Explicitly, from Eqs. \eqref{chipotential} and \eqref{SRparameters} we obtain that $U/\epsilon \propto \lambda/\xi^2$ in the metric case and $U/\epsilon \propto \lambda/\xi$ in the Palatini formulation. This parametric dependence matches the one of the running non-minimal coupling~\eqref{eq:runx}. Moreover, when expressed in terms of inflationary parameters, Eq. \eqref{eq:runx} will be identical in both the metric and Palatini formulation:
\be \label{eq:infc}
	\xi_E(\mu) 
	= \frac{(0.027)^4}{24 \pi^2 N^2} \log\bigg(\f{\mu}{\mu_0}\bigg) 
	= 6\times10^{-13} \, \log\bigg(\f{\mu}{\mu_0}\bigg)  \qquad \mbox{metric and Palatini},
\ee
where in the last line we assumed $N=60$. This equivalence is derived from the tree level inflationary predictions derived in Sec.~\ref{model}, which is justified when the effect of quantum corrections on inflation is negligible.

The smallness of the quantum corrections to inflation is already implied by Eq. \eqref{eq:infc}, but to make a direct estimate of their size let us set $\xi_E$ to zero at the end of inflation, $\xi_E(\chi=\chi_f)=0$. Then, $60$ e-folds before the end of inflation $\xi_E(\chi=\chi_i)\simeq 5\times10^{-13}$ in the metric case and $\xi_E(\chi=\chi_i)\simeq 4\times10^{-18}/\sqrt{\lambda}$ in the Palatini case. In both cases the absolute relative effect of the running $\xi_E$ on both $n_s$ and $r$ is $\lesssim10^{-10}$. These corrections can thus be safely neglected, as they are several orders of magnitude smaller than the errors induced by using the slow-roll approximation. The smallness of the quantum corrections is, however, not surprising, since quantum corrections to an exponentially flat Einstein frame potential such as~\eqref{chipotential} are expected to be small~\cite{Herranen:2015aja}.

\section{Conclusions}
\label{conclusions}

In this paper we studied quantum corrections to single field inflation non-minimally coupled to gravity. We performed a comparison between results obtained in the usual metric approach to gravity and the Palatini formalism where the connection is initially treated as a free variable. We presented a simple derivation based on first principles of the one-loop quantum effective potential on a homogeneous and isotropic background, valid for general potentials. The calculation was performed in the classical Einstein frame where the non-minimal coupling between the scalar curvature of gravity and the inflaton field is removed via a Weyl scaling of the metric. Nevertheless, as our calculation verified, in both cases the quantum corrections mix curvature and the scalar field and a non-minimal coupling is regenerated at one-loop. Specialising to quartic self-interactions we derived the leading approximation for the RG improved effective potential in both formalisms along with the respective beta-functions. It was shown that the quantum corrections to inflationary predictions are negligible, of the order of $10^{-10}$ for both cases. Specifically, the tree level large $N$ expansion~\eqref{nsr} remains intact.

Quantum corrections to inflation have been calculated in many models and approaches over the years and in this framework our results follow familiar lines: the flatness of the potential is generically not spoiled by loop corrections which can be linked to the observational fact that the inflationary perturbations are small, as is visible in (\ref{eq:infc}) and the discussion leading to it. As our calculation verified, this is true also in the Palatini formulation of gravity. However, we wish to highlight an often overlooked fact, at least perhaps in the context of Higgs inflation studies, which is that the minimally coupled form of the action will be spoiled by quantum corrections. This in fact is the dominant loop correction, which necessitates one to include background curvature in the field theory calculation. The effect is visible already at the one loop level, but becomes even more apparent in the renormalisation group approach, which was shown to lead in a well-defined effective field theory sense to a non-zero beta function for $\xi_E$, i.e. the non-minimal coupling generated in the Einstein frame. As expected this is equally true both in the metric and Palatini formulations of gravity, but the beta functions for $\xi_E$ were interestingly discovered to differ between the two approaches. However, they turn out to be identical in magnitude during inflation. One may of course perform an additional Weyl scaling to remove non-minimal terms generated via the one loop quantum corrections, but upon calculating further corrections the same procedure should then be repeated order by order to maintain a strictly minimally coupled form of the theory. The Einstein frame therefore depends on the order of perturbation theory.

Our study can be straightforwardly extended by the inclusion of additional matter fields or by considering more general modified gravity theories. Interaction between matter and the inflaton field may also generate significant contributions to the effective potential. Furthermore, unlike for scalars or gauge bosons, the fermionic kinetic terms depend on the connection and will thus contribute to the equation of motion for the connection. Thus models with fermions introduce new subtleties when the connection is promoted to be an independent dynamical variable, as is the case in the Palatini formulation. Similar features may be observed in modified gravitational theories such as $f(R)$ gravity, in which case the metric and Palatini formulation can differ significantly already at tree level~\cite{Olmo:2011uz}. It would be interesting to see their detailed effect on models of inflation.

\section*{Acknowledgements}
We thank Syksy R\"as\"anen for discussions. T.M. and T.T. are supported by the STFC grants ST/P000762/1 and ST/J001546/1, respectively. V.V. and H.V are supported by the Estonian Research Council grant IUT23-6 and the ERDF Centre of Excellence project No TK133, and H.V in addition by the Estonian Research Council via a Mobilitas Plus grant MOBTT5.

\bibliography{palatini.bib}

\end{document}